\documentclass[aps,prl,twocolumn,groupedaddress,showpacs,showkeys]{revtex4}

\usepackage{graphicx}
\usepackage{bm}
\usepackage{amsmath}

\begin{document}

\title{Coherence in Microchip Traps}

\author{Philipp~Treutlein}\email[E-mail:]{philipp.treutlein@physik.uni-muenchen.de}
\author{Peter~Hommelhoff}\altaffiliation[Present address: ]{Varian
  Physics 220, Stanford University, Stanford, CA 94305, U.S.A.}
\author{Tilo~Steinmetz} \author{Theodor~W.~H{\"a}nsch} \author{Jakob~Reichel}
\affiliation{Max-Planck-Institut f{\"u}r Quantenoptik und Sektion Physik der
Ludwig-Maximilians-Universit{\"a}t, Schellingstr.4, 80799 M{\"u}nchen, Germany}

\date{\today}

\begin{abstract}
  We report the coherent manipulation of internal states of neutral
  atoms in a magnetic microchip trap.  Coherence lifetimes exceeding
  1\,s are observed with atoms at distances of $5-130\, \mu$m from the
  microchip surface.  The coherence lifetime in the chip trap is
  independent of atom-surface distance within our measurement accuracy, and agrees well with the
  results of similar measurements in macroscopic magnetic traps.  Due
  to the absence of surface-induced decoherence, a miniaturized atomic
  clock with a relative stability in the $10^{-13}$ range can be
  realized.  For applications in quantum information processing, we
  propose to use microwave near-fields in the proximity of chip wires
  to create potentials that depend on the internal state of the atoms.
\end{abstract}


\maketitle

Magnetic microchip traps provide one of the few available techniques
for manipulating neutral atoms on the micrometer scale, and the
only technique so far that enables nonperiodic,
built-to-purpose micron-sized potentials \cite{ReichelFolman02}.  The
on-chip creation of Bose-Einstein condensates \cite{Ott01,Haensel01b}
and the highly controlled manipulation of atomic motion in `atomic
conveyor belts', waveguides, and thermal beam
splitters \cite{ReichelFolman02} are examples of the versatility of such
`atom chips'. Due to these possibilities, chip traps 
are promising candidates for the implementation of
quantum gates \cite{Calarco00}, quantum simulations
\cite{Jane03}, and interferometric sensors \cite{Kasevich02}.  The
ability to manipulate superpositions of internal states of
the trapped atoms is essential for most of these
applications. In quantum information processing (QIP), two
internal states $|0\rangle$ and $|1\rangle$ of the atom serve as qubit
states. To perform gate operations, long coherence
lifetimes of the superposition states $\alpha |0\rangle + \beta
|1\rangle$ are required and therefore decoherence
processes have to be avoided.  Atoms in chip traps, however,
can potentially suffer from a reduction of the coherence lifetime due
to interaction with the surface of the chip
\cite{Henkel03}, in addition to other decoherence mechanisms which are
also present in macroscopic traps \cite{Harber02}.

\begin{figure}[b]
\includegraphics{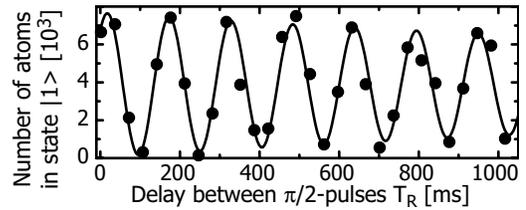}
\caption{\label{fig:RamseyTime} Ramsey spectroscopy of the $|0\rangle 
  \leftrightarrow |1\rangle$ transition with atoms held at a distance
  $d=9\,\mu$m from the chip surface. An exponentially damped sine fit
  to the Ramsey fringes yields a
  $1/e$ coherence lifetime of $\tau_c=2.8\pm 1.6$\,s.
   Each data point corresponds to a single shot of the experiment.
  }
\end{figure}

In this letter, we demonstrate coherent
manipulation of internal atomic states in a magnetic microchip trap.  We
create superpositions of two hyperfine ground states of $^{87}$Rb
atoms in a thermal ensemble close to quantum degeneracy and perform
Ramsey spectroscopy to determine the coherence lifetime
(Fig.~\ref{fig:RamseyTime}).  With atoms at distances of
$5-130\,\mu$m from the surface of the chip, we observe coherence
lifetimes exceeding 1\,s. These lifetimes
are independent of the atom-surface distance, and agree well
with those observed in macroscopic magnetic traps
\cite{Harber02}.

The observed robustness of the superposition states is an extremely
encouraging result for atom chip applications in QIP
and opens a new perspective on applications in precision
metrology. We demonstrate an
atomic clock in the chip trap and measure the relative stability of
its transition frequency.
 Our measurements show that a portable
atom chip clock with a relative stability in the 
$10^{-13}\tau^{-1/2}/\sqrt{\textrm{Hz}}$ range is a realistic goal.

To realize the collisional phase gate proposed in
\cite{Calarco00}, a state-selective potential is
needed. We point out that state selectivity for our
 state pair can be provided
by microwave potentials. These potentials, considered in the
early 90s \cite{Agosta89,Spreeuw94} but abandoned later,
 gain new actuality as near-field traps on atom chips.


To achieve long coherence lifetimes with magnetically trapped atoms in the proximity of the chip surface,
we choose the $|F=1,m_F=-1\rangle \equiv |0\rangle$ and
$|F=2,m_F=1\rangle \equiv |1\rangle$ hyperfine levels of the
$5S_{1/2}$ ground state of $^{87}$Rb. 
The magnetic moments
of the two states are approximately equal. 
At a magnetic field of $B_0 \sim 3.23$\,G, both states
experience the same first-order Zeeman
shift and the remaining magnetic field dependence of the transition frequency $\nu_{10}$ is minimized
\cite{Harber02}. In all of our experiments, we therefore adjust the
field in the center of the trap to $B_0$.
This greatly reduces spatial inhomogeneities of $\nu_{10}$ in the
 trap, since both states experience the same
 potential to good approximation. Furthermore, superpositions of this state pair are
particularly robust against decoherence due to magnetic field
noise. Thermal magnetic near-field noise has been predicted to be a relevant atom-surface coupling mechanism in the regime of distances studied here \cite{Henkel03}. For our state pair, this source of decoherence is suppressed by more than $10^6$ (the squared ratio of electron and nuclear magnetic moment) and we expect surface decoherence rates to be negligible.

\begin{figure}[t]
\includegraphics{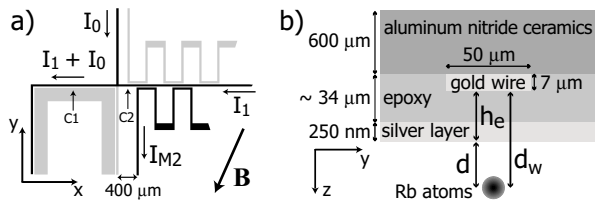}
\caption{\label{fig:Substrate} (a) Layout of the relevant wires on the 
  chip substrate. C1: position of the initial magnetic
  trap. C2: position of the `measurement trap' used in the experiments. 
  (b) Layer structure of the substrate. Current is carried by gold wires $h_e \sim 27\,\mu$m below the silver surface. $d$ denotes the atom-surface distance, while $d_w$ refers to the atom-wire distance.
  }
\end{figure}

Our experimental setup has been previously described
\cite{Reichel99,Haensel01b}. The magnetic
potentials are created by sending currents through microscopic
gold conductor patterns on a substrate and
superposing a homogeneous bias field
(Fig.~\ref{fig:Substrate}).  The preparation of the atomic ensemble
proceeds in a multi-step sequence involving loading of the microtrap
from a mirror-MOT, compression of the trap and evaporative cooling
\cite{Haensel01b}. By the end of this sequence, a
Ioffe-type `measurement trap' centered at the position C2 in
Fig.~\ref{fig:Substrate}a contains a thermal atomic ensemble of
typically $1.5 \times 10^4$ atoms in state $|0\rangle$ at a
temperature of $0.6\,\mu$K.
The measurement trap is created by the currents $I_0$, $I_1$, and
$I_{M2}$ and the bias field $\mathbf{B}=(B_x,B_y,0)$ shown in
Fig.~\ref{fig:Substrate}a. By adjusting all three currents and $B_y$,
the atoms can be placed at distances $d=0-130\,\mu$m from the chip
surface (see Fig.~\ref{fig:Substrate}b) with only small changes in
the shape of the magnetic potential. For each
experimentally studied distance, the field $\mathbf{B}_\textrm{min}$
at the center of
the trap was calibrated spectroscopically
\cite{Harber02} and set to $|\mathbf{B}_\textrm{min}|=B_0$ by
adjusting $B_x$. Typical experimental
parameters are $I_0=500$\,mA, $I_1=120$\,mA, $I_{M2}=700$\,mA,
$B_y=-5.50$\,G, and $B_x=-2.18$\,G, leading to trap frequencies
$(f_x,f_y,f_z)=(50,350,410)$\,Hz at $d=9\,\mu$m. 
The atoms are held in the measurement trap while the
coherent internal state manipulation is performed. After the
manipulation, the trap is switched off within $150\,\mu$s
and the atoms are detected after a time of flight of typically 4\,ms.
Atoms are detected by absorption imaging
on the $F=2\rightarrow F^\prime=3$ transition. This allows direct
determination of the number of atoms in state $|1\rangle$, $N_1$. To
alternatively determine the number of atoms in state $|0\rangle$,
$N_0$, we first blow away all atoms in $|1\rangle$ with the resonant
probe light. The $|0\rangle$ atoms are then optically pumped to
$|F=2,m_F=+2\rangle$ and imaged as before.

\begin{figure}[t]
\includegraphics{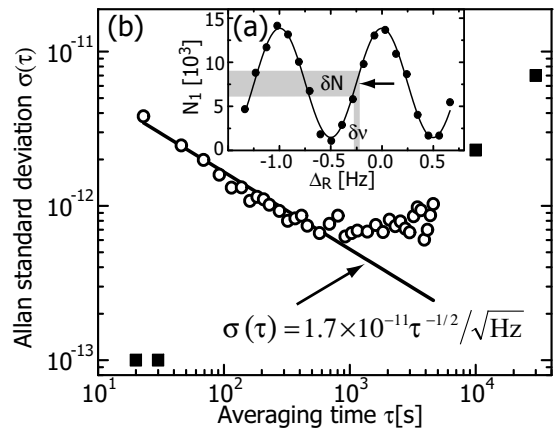}
\caption{\label{fig:Allan} (a) Ramsey resonance for $T_R=1$\,s in the microtrap at $d=54\,\mu$m
  from the surface (solid circles). The line is a sinusoidal fit to
  the data. The arrow indicates the operating point on the slope of
  the resonance used in the measurement of $\sigma(\tau)$. Frequency
  fluctuations $\delta \nu$ lead to fluctuations $\delta N$ in the
  detected number of atoms. (b) Measured Allan standard deviation 
  $\sigma(\tau)$ of the atomic clock in the microtrap compared to the
  quartz reference oscillator (open circles). The solid line is a fit with
  $\sigma(\tau)\propto \tau^{-1/2}$ representing the performance of
  the atomic clock. For $\tau>6\times 10^2$\,s the drift of the
  quartz reference becomes apparent. The manufacturers'
  specification of the quartz stability is shown in solid squares.}
\end{figure}

Coherent internal state manipulation is achieved by coupling 
$|0\rangle$ and $|1\rangle$ through a two-photon microwave-radio frequency
transition. The microwave frequency $\nu_\textrm{mw}$ is detuned
$1.2$\,MHz above the $|F=2,m_F=0\rangle$ intermediate state and
radiated from a sawed-off waveguide outside the vacuum chamber. The
radio frequency $\nu_\textrm{rf}$ is either applied to the same
external coil that is used for evaporative cooling, or to
a wire on the chip.
$\nu_\textrm{mw}$ and $\nu_\textrm{rf}$ are phase locked to an ultrastable 10\,MHz
quartz oscillator (Oscilloquartz OCXO 8607-BM,
 cf. Fig.~\ref{fig:Allan}). By applying the two-photon drive for a variable
time and detecting the number of atoms transferred to
$|1\rangle$, we observe Rabi oscillations with a resonant two-photon
Rabi frequency of $\sim 0.5$\,kHz. In this way, single-qubit rotations
can be realized. The $\pi$-pulse transition probability 
is $N_1/(N_0 + N_1)=95\pm 5$\,\%.

To test for decoherence of the superposition states, we perform Ramsey
spectroscopy:
The atoms in state $|0\rangle$ are held in the measurement trap for a
time $T_H$ before a first $\pi/2$-pulse creates a 
superposition of $|0\rangle$ and $|1\rangle$. After a 
delay $T_R$, a second $\pi/2$-pulse is applied, and the resulting
state is probed. Time-domain Ramsey fringes are recorded by varying $T_R$ while keeping
$\Delta_R=\nu_\textrm{mw}+\nu_\textrm{rf} - \nu_{10}$
fixed ($\Delta_R\ll \nu_{10}\simeq 6.8$\,GHz). 
Alternatively, frequency-domain Ramsey fringes are recorded by scanning $\Delta_R$
with constant $T_R$. 
Loss of coherence can show up in different ways in the Ramsey signal. A spatial
variation of $\nu_{10}$ across the atomic ensemble leads to a decay of
the fringe contrast, while temporal fluctuations of
$\nu_{10}$ lead to increasing phase noise as $T_R$ is increased.


Figure~\ref{fig:RamseyTime} shows time-domain Ramsey fringes. 
The number of atoms detected in state $|1\rangle$ oscillates
at the frequency difference $\Delta_R=6.4$\,Hz, while the interference
contrast decays with a coherence lifetime of $\tau_c=2.8\pm 1.6$\,s.
The measurement of Fig.~\ref{fig:RamseyTime} was performed
at $d=9\,\mu$m from the room-temperature chip
surface. In \cite{Harber02}, similar coherence lifetimes are reported
for the same state pair, but with
atoms in a macroscopic magnetic trap, far away from any material
objects. This suggests that atom-surface
interactions indeed do not limit the coherence lifetime in our 
experiment.

\begin{figure}[t]
  \includegraphics{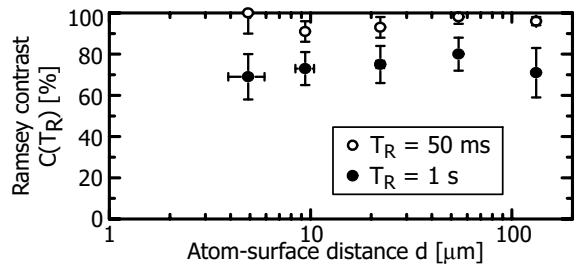}
\caption{\label{fig:ContrastDist} Contrast $C(T_R)$ of the Ramsey
  fringes as a function of atom-surface distance $d$ for two values of
  the time delay $T_R$ between the $\pi/2$-pulses. For each data point,
  $C(T_R)=(N_\textrm{max}-N_\textrm{min})/(N_\textrm{max}+N_\textrm{min})$
  was obtained from a sinusoidal fit to frequency-domain Ramsey
  fringes. $N_\textrm{max}$ ($N_\textrm{min}$) is the maximum
  (minimum) of the oscillation in $N_1$. The data points for $d=(5,9,22,54,132)\,\mu$m were measured with atomic ensembles of temperatures $T=(0.2,0.6,0.7,0.6,0.3)\mu$K and peak densities $n_0=(4,3,1,1,5)\times 10^{12}$\,cm$^{-3}$.
}
\end{figure}

To further probe for surface effects, we study decoherence 
as a function of atom-surface distance $d$ (Fig.~\ref{fig:ContrastDist}). 
At each distance, we record frequency-domain Ramsey oscillations for
several values of $T_R$ and determine the contrast $C(T_R)$ of each
oscillation. Figure~\ref{fig:ContrastDist} shows the result 
for $T_R=50$\,ms and $T_R=1$\,s. Within the experimental
error, the contrast does not show a dependence on atom-surface
distance.
Additionally, we have compared the
signal-to-noise ratio $S/N$ of the interference signals at different $d$. We typically observe $S/N=6$ for $T_R=1$\,s, where $S$ is the peak-to-peak amplitude of the sinusoidal fit to the Ramsey oscillation and
$N$ is the standard deviation of the fit residuals over one
oscillation period. $S/N$ is independent of $d$ within experimental error, indicating that the
processes causing amplitude and phase fluctuations of the interference
signal do not depend on atom-surface distance on this time scale.

The observed decoherence is mainly due to a
combination of the residual differential Zeeman shift and the 
density-dependent collisional shift of $\nu_{10}$ across the ensemble
\cite{Harber02}. Consequently, we observe a dependence
of $\tau_c$ on the temperature $T$ and on the peak density
$n_0$ of the atoms. To avoid systematic errors, 
we have checked that there is no systematic variation of
$T$ and $n_0$ as $d$ is varied (see Fig.~\ref{fig:ContrastDist}). The observed noise on
the Ramsey oscillation is mostly phase noise and can be attributed to
ambient magnetic field fluctuations (see below). For $T_R>1$\,s the
phase noise increases and obscures the oscillation even before
the contrast has completely vanished.

To calibrate the atom-surface distance $d=d_w-h_e$ (see Fig.~\ref{fig:Substrate}b), we use two
methods: For $d>10\,\mu$m, we apply a technique described in
\cite{Schneider03} in which the imaging beam 
is slightly tilted towards the chip surface.
For $d<10\,\mu$m, the distance $d$ can no longer be resolved by our imaging system.
In this case, $d_w$ can be determined from a simulation of the trapping potential, but $d$ is affected by uncertainties in $h_e$, which was not well-controlled during manufacture. To determine $h_e$, we move the atoms close to the surface and measure the remaining atom fraction after 10\,ms as a function of $d_w$ \cite{Lin03}. 
Using a model for atom loss due to the attractive Casimir-Polder surface potential \cite{Lin03}, which contains $h_e$ as the only free parameter, we determine $h_e=27.1\,\mu$m with a statistical error of $0.1\,\mu$m. 
Including errors in the model parameters, such as trap frequencies and temperature, we estimate an uncertainty in $d$ of $\pm 1\,\mu$m (error bars in Fig.~\ref{fig:ContrastDist}).

Thermal magnetic field noise driving spin-flip transitions to
untrapped states \cite{Henkel03} and surface evaporation have been 
observed to limit the lifetime $\tau_N$ of the atomic \emph{population} 
near a surface \cite{Jones03,Harber03,Lin03}. We also observe
these effects. The trap lifetime in state
$|0\rangle$ decreases from $\tau_N=11$\,s for $d>20\,\mu$m to
$\tau_N=1.6$\,s for $d=5\,\mu$m 
($\tau_N$ for state $|1\rangle$ is slightly lower due to stronger coupling to the surface \cite{Henkel03} and dipolar relaxation). 
For $d< 5\,\mu$m, this atom loss prohibits coherence measurements with $T_R\sim 1$\,s.
To distinguish loss of population from loss of
\emph{coherence}, $T_H + T_R$ is kept constant during the Ramsey scan
by appropriately adjusting the hold time $T_H$ for each value of
$T_R$. Thus, the overall time the atoms spend close to the
surface is independent of $T_R$. $N_0+N_1$ at the time
of detection therefore remains approximately constant.
The small lifetime difference of the two states reduces $C(T_R=1\,\textrm{s})$ by less than $5$\,\%.


One motivation for atom chip research is the perspective of
creating miniaturized cold-atom devices. Due to the
long coherence lifetime, it is natural to consider utilizing
the $|0\rangle \leftrightarrow |1\rangle$ transition
in an atomic clock on the chip.  We demonstrate the
principle of such a clock and measure its frequency stability relative
to the quartz reference oscillator. 
Figure~\ref{fig:Allan}a shows frequency-domain Ramsey fringes
for $T_R=1$\,s. We set the two-photon drive to
the slope of the Ramsey resonance (arrow in the
figure), and repeat the experiment many times with a cycle period of
23\,s. Any temporal drift $\delta \nu$ of
$\nu_{10}$ with respect to the reference will
change $\Delta_R$ and therefore show up as a variation $\delta N$ of
$N_1$.
From repeated measurements of $\delta N$ we determine the relative
frequency fluctuations $\delta \nu / \nu_{10}$. In
Fig.~\ref{fig:Allan}b we plot the Allan standard deviation
$\sigma(\tau)$ \cite{Santarelli99} of $\delta \nu / \nu_{10}$ as a
function of averaging time $\tau$.
For short $\tau$, 
$\sigma(\tau)$ decreases as
$\sigma(\tau)=1.7\times 10^{-11} \tau^{-1/2}/\sqrt{\textrm{Hz}}$,
corresponding to shot-to-shot fluctuations of $\delta \nu=24$\,mHz r.m.s.
For $\tau>6\times 10^2$\,s, the long-term drift of
the reference leads to a departure from the $\tau^{-1/2}$
line.  We have modeled the frequency fluctuations 
and can account for the observed value of $\delta \nu$. It is
dominated by ambient magnetic field fluctuations of $\sim 5$\,mG. 
Smaller contributions are due to $\sim 4$\,\% variations 
in total atom number leading to variations of the
 collisional shift \cite{Harber02}, and due to
imperfections of the detection system. We estimate that
realistic improvements --- magnetic shielding, operation in a 
shallower trap at lower atomic density, shot-noise limited detection, and a 6\,s cycle (which is realistic in a chip trap \cite{Haensel01b}) --- will lead to a frequency stability in the 
$10^{-13}\tau^{-1/2}/\sqrt{\textrm{Hz}}$ range. While this does not reach the
stability level of fountain clocks, a chip-based clock has the 
advantage of a simple, compact and portable setup.


Applications of magnetic microtraps in QIP
require long coherence lifetimes of the qubit in the presence of unavoidable
magnetic field noise. A state pair with equal magnetic moments
is therefore much better
suited than any other combination of ground state
sublevels.  In \cite{Calarco00}, a phase gate with atoms in a
magnetic microtrap was proposed, and a gate time of $0.4$\,ms
was estimated. 
Implementing this gate with atoms in states $\left\{ |0\rangle,|1\rangle\right\}$ located $\sim 5\,\mu$m above a chip with micron-sized wires, $\sim 10^3$ gate operations could be performed before decoherence from magnetic noise occurs.
The gate requires state-dependent potentials. However, a combination of
static magnetic and electric fields, as considered in
\cite{Calarco00,Krueger03}, does not provide state-selectivity
for our state pair, whose magnetic
moments and electrostatic polarizabilities are equal.  Instead, we
propose to apply tailored microwave near-fields and make use of the AC
Zeeman effect (the magnetic analog of the AC Stark effect).
In $^{87}$Rb, AC Zeeman potentials derive from magnetic dipole transitions 
near $\omega_0/2\pi= 6.835$\,GHz between the $F=1$ and $F=2$
hyperfine manifolds of the ground state.
The magnetic component of a microwave field of frequency $\omega_0+\Delta$ couples the $|F=1,m_F\rangle$ to the $|F=2,{m_F}^\prime\rangle$ sublevels and leads to energy shifts that depend on $m_F$ and ${m_F}^\prime$.
In a spatially varying microwave field, this results in a state-dependent potential landscape.

A microwave trap based on AC Zeeman potentials was
proposed in \cite{Agosta89} and experimentally demonstrated in
\cite{Spreeuw94}. This trap employed microwave radiation in the
far field of the source. Due to the centimeter wavelength
$\lambda_\textrm{mw}$ of the radiation, field gradients were
weak and structuring the potential on the micrometer scale is
impossible. In a chip trap, on the other hand, atoms are
trapped at distances $d\ll \lambda_\textrm{mw}$ from the chip wires. 
Thus, they can be manipulated with microwave near fields, generated by
microwave currents in the wires, which may be fed from
a stripline \cite{Edwards81}. 
 In the near field of the
currents, the magnetic component of the microwave field
has the same position dependence as a static magnetic field created by
equivalent dc currents. In this way, state-dependent AC Zeeman
potentials varying on the micrometer scale can be created. 
In combination with state-independent static magnetic traps, the
 potential geometries required for QIP can be
realized.

To be specific, we consider a static-field trap at $d=10\,\mu$m from an additional chip wire carrying a microwave current of 20\,mA$_\textrm{pp}$. 
The wire is oriented such that the magnetic component of the microwave field at the position of the atoms is polarized parallel to the local static magnetic field. 
The microwave couples $|0\rangle \leftrightarrow |F=2,m_F=-1\rangle$ and $|F=1,m_F=1\rangle \leftrightarrow |1\rangle$ with identical resonant Rabi frequencies $\Omega_R/2\pi=2.4$\,MHz.
The Zeeman splitting due to the static field (a few MHz) prevents two-photon transitions to other sublevels driven by polarization impurities. 
For $\Delta/2\pi=50$\,MHz, $\Omega_R \ll \Delta$ and the coupling changes the static magnetic moment of the qubit states by only $\sim 10^{-4} \mu_\mathrm{B}$ such that both states still experience the same static-field potentials. 
The microwave, on the other hand, leads to a differential energy shift of $|0\rangle$ and $|1\rangle$, $U_\textrm{mw}\simeq \hbar \Omega_R^2/2\Delta = h \cdot 58$\,kHz, sufficiently large for the state-selective manipulation required in QIP.
Besides this application, AC Zeeman
potentials can be used to create microtraps for atoms in
hyperfine sublevels such as $m_F=0$, which cannot be trapped in a
static magnetic trap.


In conclusion, we have performed coherent internal state manipulation
in a magnetic microchip trap with coherence lifetimes exceeding 1\,s at
distances down to $5\,\mu$m from the chip surface. This paves the way
for a variety of applications, most notably
chip-based quantum gates and atomic clocks.

We thank T.~Udem and M.~Zimmermann for the crystal reference oscillator. Work supported in part by the EU's IST program (ACQP, IST-2001-38863).

\end{document}